\documentclass[12pt,british]{revtex4}
\usepackage[T1]{fontenc}
\usepackage[latin9]{inputenc}
\usepackage[letterpaper]{geometry}
\usepackage{color}
\usepackage{amsmath}
\usepackage{graphicx}
\usepackage{amssymb}

\makeatletter
\@ifundefined{textcolor}{}
{%
 \definecolor{BLACK}{gray}{0}
 \definecolor{WHITE}{gray}{1}
 \definecolor{RED}{rgb}{1,0,0}
 \definecolor{GREEN}{rgb}{0,1,0}
 \definecolor{BLUE}{rgb}{0,0,1}
 \definecolor{CYAN}{cmyk}{1,0,0,0}
 \definecolor{MAGENTA}{cmyk}{0,1,0,0}
 \definecolor{YELLOW}{cmyk}{0,0,1,0}
 }

\@ifundefined{definecolor}
 {\usepackage{color}}{}
\makeatother

\makeatother

\usepackage{babel}

\begin{document}

\title{Configurational Prigogine-Defay ratio}

\author{J.-L. Garden}

\email{jean-luc.garden@grenoble.cnrs.fr}

\affiliation{Institut Néel, CNRS et Université Joseph Fourier, PB 166, 38042 Grenoble
Cedex 9, France}

\author{H. Guillou}

\affiliation{Institut Néel, CNRS et Université Joseph Fourier, PB 166, 38042 Grenoble
Cedex 9, France}

\author{J. Richard}

\affiliation{Institut Néel, CNRS et Université Joseph Fourier, PB 166, 38042 Grenoble
Cedex 9, France}

\author{L. Wondraczek}

\affiliation{Departement of Materials Science, Chair of Glass and Ceramics, University
of Erlangen-Nuernberg, 91058 Erlangen, Germany}
\begin{abstract}
Classically, the Prigogine-Defay (PD) ratio involves differences in
isobaric volumic specific heat, isothermal compressibility and isobaric
thermal expansion coefficient between a super-cooled liquid and the
corresponding glass at the glass transition. However, determining
such differences by extrapolation of coefficients that have been measured
for super-cooled liquid and glassy state, respectively, poses the
problem that it does not take into account the non-equilibrium character
of the glass transition. In this paper, we asses this old question
by taking into account the gradual change of configurational contributions
to the three thermodynamic coefficients upon varying temperature and
pressure. Macroscopic non-equilibrium thermodynamics is applied to
obtain a generalized form of the PD ratio. The classical PD ratio
can then be taken as a particular case of this generalization. Under
some assumptions, a configurational PD ratio (CPD ratio) can be expressed
in terms of fictive temperature and fictive pressure what, hence,
provides the possibility to experimentally verify this formalism.
Noteworthy and differing from previous approaches towards the PD ratio,
here, the glass transition is considered as non-isoaffine.

\clearpage{} 
\end{abstract}
\maketitle

\section{Introduction}

Upon vitrification, some of the thermodynamic properties of a glass
former undergo a jump which can be observed experimentally. Classically,
the Prigogine-Defay (PD) ratio was defined as the ratio of the jumps
of the following thermodynamic quantities:\begin{equation}
\Pi=\frac{1}{VT_{g}}\left\{ \frac{\triangle C_{p}\triangle\kappa_{T}}{\left(\triangle\alpha_{p}\right)^{2}}\right\} _{T=T_{g}}\label{eq:PD 1}\end{equation}
 where $\triangle C_{p}$, $\triangle\kappa_{T}$ and $\triangle\alpha_{p}$
are the differences taken at the glass temperature, $T_{g}$ , of
the extrapolated measured isobaric heat capacity, isothermal compressibility
and isobaric thermal expansion coefficient as observed in liquid and
glassy states, respectively. $V$ is the volume of the system. The
PD ratio has initially been derived by Prigogine and Defay, using
classical thermodynamics of irreversible processes and De Donder's
order parameter framework \cite{prigo}. In their original definition,
the sign {}``$\Delta$'' represents a difference between a state
at constant affinity (isoaffine state) and a state at constant order
parameter (that is, the glassy state). Indeed, since the affinity
and the order parameter are thermodynamically conjugated, in analogy
to the well-known expression of the difference between isobaric and
isochoric heat capacity ($C_{p}-C_{V}),$ an expression of the difference
between isoaffine and iso-order-parameter heat capacity ($C_{A}-C_{\xi})$
can be derived. So, in allowing the affinity to tend towards zero,
the sign {}``$\triangle$'' represents the super-cooled liquid-to-glassy
difference and, hence, the common definition of the PD ratio is obtained.
If applied to second order phase transitions, simple equilibrium thermodynamic
arguments show that this quantity is equal to unity. This can be demonstrated
by expressing the continuity of volume and entropy (first order derivatives
of the Gibbs free energy) following Ehrenfest's classification, when
the system undergoes a phase change \cite{mckenna1}.

For the case of glass transitions, however, numerous experiments have
revealed that this ratio may differ from unity. That is, in some cases,
PD < 1 has been observed, but more often, experimental value of PD
> 1 are obtained (as an extreme example, for the case of vitreous
silica, classical experimental approaches yield a value of PD > 10000).
In a pioneering study, Davies and Jones were transferring De Donder's
nomenclature to the classical field of glass science\cite{davies}.
They interpreted the experimental observation of PD$\neq$1 as an
evidence that multiple (more than one) order parameters must be introduced
to completely describe the glass transition in a real system. Over
the following decades, the so-called {}``order parameter approach''
has then been applied and refined by Moynihan, DiMarzio, Goldstein,
Gupta and others \cite{dimarzio,goldstein,gupta,berg,lesikar} and
became widely accepted. A more general approach towards the PD ratio
was undertaken by Lesikar and Moynihan. In their consideration, the
value of the PD ratio results from a thermodynamic stability condition,
and can be generalized to unconventional thermodynamic parameters
such as electric polarisation \cite{lesikar2}. It was assumed that
each thermodynamic variable is governed by its corresponding order
parameter, what generally implies PD > 1. They also have derived a
particular condition for which the PD ratio is equal to unity even
if several order parameters are taking part in the glass transition
\cite{lesikar2}. The topic has been reviewed in a very complete form
by Nemilov \cite{nemilov}. In further studies, also, Gutzow and Schmelzer
have discussed the PD ratio mostly from viewpoint of its genuine definition
\cite{gutzow}.

During the last decade, however, with the publication of several new
papers, the debate has been reopened. Similar to Ritland's original
concept of the fictive temperature \cite{ritland1,ritland2}, a phenomenological
{}``two temperature thermodynamic approach'' was developed by Nieuwenhuizen
in which he reconsiders various classical problems of the glass transition
\cite{Nieu2,Nieu}. Among them is counted the non-unity of the PD
ratio. It was shown that by taking into account the configurational
entropy of a supercooled system, one of the two Ehrenfest relations
is modified what, consequently yields PD $\neq$ 1. The problem was
approach from another angle by Ellegaard and al. \cite{dyre1}. In
their study, they are introducing a dynamic PD ratio where only the
imaginary parts of the involved complex thermoviscoelastic coefficients
are taken into account. They demonstrate rigorously that under such
assumptions PD may become unity at all frequencies even when only
a single order parameter is taken into account. Experimental and theoretical
works have confirmed this for volume-enthalpy correlated glass-formers
\cite{dyre2,pick}. Eventually, in 2006, Schmelzer and Gutzow, in
re-formulating a thermodynamic description by means of classical thermodynamics
of irreversible processes, derived a new expression of the PD ratio
\cite{gutzow2}. A new expression of the P.D. ratio was derived. Their
expression highlights the role of thermodynamic affinity, explicitly
stating that the considered system is out of thermodynamic equilibrium,
and that the glass transition does not follow Ehrenfest's original
definition of a phase transition (where a discontinuity in some derivative
of Gibbs free energy occurs at a precise temperature or pressure).
More precisely, the transition process takes place over a definite
temperature or pressure range. Within this range, the system starts
to deviate from thermodynamic equilibrium. Thus, in their formulation,
the difference {}``$\triangle$'' attains its genuine formulation:
a difference between a state at constant affinity and a state at constant
order parameter. In the present study, based on these considerations,
a simple expression of the PD ratio is derived, based solely on enthalpy
and affinity. Subsequently, it is demonstrated within this formulation,
that PD can differ from unity, independently on the number of involved
order parameters.

Under the {}``single order parameter assumption'', we provide a
generalization of Schmelzer and Gutzow's expression of the PD ratio,
and we extend the approach to other than isoaffine transformations.
The configurational parts of the three thermodynamic coefficients,$C_{p}$,
$\kappa_{T}$ and $\alpha_{p}$ are used directly in order to define
a configurational Prigogine-Defay (CPD) ratio. The classical expression
of the PD ratio as well as the expression of Schmelzer and Gutzow
become particular cases of this definition, where the creation of
affinity occurs as a direct result of changes in temperature or pressure.
This demonstrates the complexity of the phenomenological behaviour
of the glass transition as a thermodynamically irreversible process.

The paper is written as follows: In section 2, the non-equilibrium
thermodynamic approach of the glass transition is briefly exposed.
In section 3, the CPD ratio is proposed. A tri-dimensional diagram
is then depicted and the glass transition is seen as a path in this
diagram. A simple expression of the CPD ratio is then provided, revealing
only the isobaric heat capacity and the isothermal compressibility
coefficient. Then, this last expression is simply transformed to make
apparent the fictive temperature and the fictive pressure. In section
4, the unity of the CPD ratio is discussed.

\section{Non-equilibrium thermodynamic approach of the glass transition}

\subsection{The order parameter model}

A physical system at thermodynamic equilibrium is characterized by
variables or parameters defining the thermodynamic state of the system.
These variables $\left\{ x_{i}\right\} $ obey an equation of state.
This equation of state connect them together. Generally, it is difficult
to attribute precise variables to a thermodynamic system under observation,
and sometimes the choice seems subjective. From De Donder's work on
irreversible chemical reactions, it is known that a supplementary
thermodynamic variable is necessary to completely describe, thermodynamically,
a system which has been brought out of equilibrium \cite{dedonder,dedonder2}.
This variable, denoted $\xi$, was called the degree of advance of
the chemical reaction. Subsequently, it has been generalized by Prigogine,
Defay and van Rysselberghe to any irreversible process. In this generalization,
$\xi$ characterizes the advance of the process. This approach has
been used until the emergence of the thermodynamic concept of internal
variables \cite{maugin,lion}. While in chemical reactions, $\xi$
can be well-defined, in other irreversible processes, its nature is
more difficult to precise. In the glass community, $\xi$ has been
called the degree of order or the order parameter%
\footnote{Certainly this appellation comes from Landau general theory on phase
transitions, where order parameter and degree of order have been defined.
The order parameter being a particular case of degree of order for
continuous transition for which third order terms in the series expansion
of the Gibbs free energy does not appear (pure second order phase
transition). In our case, for the sake of simplicity we call the variable
$\xi$ the order parameter, despite that for glass transition the
term degree of order, or degree of advance of the process (cf. De
Donder) must be more rigorously used.%
}. Following De Donder, the variation of entropy of a system undergoing
a thermodynamic transformation is given by:\begin{equation}
dS=d_{e}S+d_{i}S\end{equation}
 where $d_{e}S$ is the reversible exchange of entropy between the
system and its surrounding (exchange of heat, work or matter) and
$d_{i}S$ is the positive generation of entropy produced within the
system itself when it is relaxing towards its equilibrium state. The
entropy production term and the order parameter are connected by the
following equality:\begin{equation}
d_{i}S=\frac{Ad\xi}{T}\geq0\end{equation}
 $T$ is the temperature of the system and $A$ the affinity of the
process. $A$ is the intensive conjugated thermodynamic variable associated
to the order parameter $\xi$. In the $\left\{ p,T\right\} $ state
ensemble, the affinity is defined as:\begin{equation}
A=-\left(\frac{\partial G}{\partial\xi}\right)_{p,T}\end{equation}
 where $G\mbox{ }$is the Gibbs free energy of the system. The affinity
is the thermodynamic force of the process exactly like gradients of
intensive variables for other processes. From the definition of the
state function $G$ ($G=H-TS$), the Berthelot-De Donder formula is
obtained:\begin{equation}
A=T\left(\frac{\partial S}{\partial\xi}\right)_{p,T}-\left(\frac{\partial H}{\partial\xi}\right)_{p,T}\label{eq:Berth de donder}\end{equation}
 where $H$ is the enthalpy. A system out of thermodynamic equilibrium
is consequently characterized by the variables $\left\{ p,T,\xi\right\} $
or $\left\{ p,T,A\right\} $and not only by $\left\{ p,T\right\} $
variables. The order parameter obviously has an equilibrium value
for any values of the couple $\left\{ p,T\right\} $. In this case,
this equilibrium value $\xi_{eq}(p,T)$, defines the condition of
thermodynamic equilibrium for the system with the necessary and sufficient
condition that $A=0$ at any instant. But in the case of the glass
transition, if the system is (for a certain time) brought out of thermodynamic
equilibrium, the three latter variables must completely define the
instantaneous state of the system. The system is not anymore constrained
to remain in the $\left\{ p,T\right\} $ equilibrium plane. Instead,
it leaves this plane to follow the value of the affinity and order
parameter, which become now time-dependent. $\xi$ and $A$ depend
of time and of the instantaneous time course of $p$ or $T$. For
example, if the system is suddenly brought out of thermodynamic equilibrium
by means of quick pressure or temperature jumps then after the arrest
of these perturbations, the variables $\xi$ and $A$ continue to
evolve until they reach their equilibrium values ($\xi_{eq}$ for
the order parameter and 0 for the affinity). This time evolution is
expressed in the physical ageing process of glasses. A glass is a
system for which the order parameter has not any possibilities to
reach such equilibrium values, because experimentally the system is
brought in a range where the relaxation time becomes too high and
$\xi$ is frozen. During the vitrification process, the heat capacity
and other thermodynamic coefficients become the sum of two separated
contributions due to time-scale separation. The first contribution
is concerned by the rapid degrees of freedom (phonon's bath) as compared
to the speed of variation of the temperature. The internal exchanges
of energy which follow a temperature perturbation can occur between
these rapid modes. In other words, the modes of vibration are thermalized.
They define the temperature $T$ of the system. The other heat capacity
contribution involves the slow internal degrees of freedom as compared
to the speed of variation of the temperature (it is assumed that these
are connected to the configurational changes inside the system). They
are driven by the variable $\xi$, and this heat capacity contribution
is called the configurational contribution%
\footnote{These two contributions are experimentally accessible with difficulties,
and some sub-decompositions can take place (see for example references
\cite{johari1,johari2}) %
}:\begin{equation}
C_{p}=\left(\frac{dH}{dT}\right)_{p}=\left(\frac{\partial H}{\partial T}\right)_{p,\xi}+\left(\frac{\partial H}{\partial\xi}\right)_{p,T}\left(\frac{d\xi}{dT}\right)_{p}=C_{p,\xi}+C_{p}^{conf}\label{eq:Cp}\end{equation}
 Upon this decomposition, the glass transition is the thermodynamic
transformation represented by the gradual freezing of these configurational
modes when the pressure or the temperature vary. That is to say that
the second contribution of the right hand side of Eq. (\ref{eq:Cp})
is progressively lost during the time course of $p$ or $T$. The
adjective {}``progressively'' or {}``gradual'' means that this
loss of configurational contribution occurs in given pressure or temperature
intervals, and not only at a definite temperature or pressure (like
in a phase transition) \cite{gutzow2,gutzow3}. \textcolor{black}{If
the principal characteristic of the glass transition is a continuous
freezing of the configurational degrees of freedom, then the variable
$\xi$ which is associated to these modes is progressively slow down
until total arrest.} In this case, the velocity of $\xi$ is completely
neglected over the temperature time rate ($\left(\frac{d\xi}{dT}\right)_{p}=0$
and thus $C_{p}^{conf}=0$). Consequently, a glassy state is a state
defined by a heat capacity (idem for the other coefficients) at constant
order parameter $C_{\xi}=\left(\frac{\partial H}{\partial T}\right)_{\xi}$
where only the phonon contribution is measured. Contrary, the heat
capacity of a liquid or a supercooled liquid contains, in addition
to the latter, the order parameter contribution which in this case
is maximum (\$C\_\{p\}\textasciicircum{}\{conf\}=Max\$). In this case,
the internal exchange of energy that follows the temperature variation
allows the system to adapt its structure to the new temperature. This
maximal contribution for each thermodynamic coefficient can be exactly
calculated as a function of the thermodynamic derivatives such as
\$(see following sections). We will see that these maximal contributions,
that are also named \$ are those that in fact are computed in the
expression of the classical PD ratio.

\subsection{Working assumptions: single order parameter, single relaxation time
and linearity}

Here we briefly asses the important assumptions used in the following: 
\begin{itemize}
\item Firstly, we suppose that one single order parameter is sufficient
to explain most of the basic features of the glass transition. An
important consequence of such an assumption is that the configurational
contributions of the three thermodynamic coefficients are all driven
by the same order parameter $\xi$:\begin{equation}
C_{p}^{conf}=\left(\frac{\partial H}{\partial\xi}\right)_{p,T}\left(\frac{d\xi}{dT}\right)_{p}\label{eq:deltaCp}\end{equation}
 \begin{equation}
\kappa_{T}^{conf}=-\frac{1}{V}\left(\frac{\partial V}{\partial\xi}\right)_{T,p}\left(\frac{d\xi}{dp}\right)_{T}\label{eq:deltaKappa}\end{equation}
 \begin{equation}
\alpha_{p}^{conf}=\frac{1}{V}\left(\frac{\partial V}{\partial\xi}\right)_{p,T}\left(\frac{d\xi}{dT}\right)_{p}\label{eq:delta alpha}\end{equation}
 Let us also indicate that in the following, others relaxational effects
are not taken into account. Secondary $\beta$-relaxation processes
or anharmonicity effects are not tackled because they are not accessible
by $C_{p}$, $\kappa_{T}$ and $\alpha_{p}$ measurements. 
\item The second assumption requests that the order parameter relaxes with
one well-defined relaxation time $\tau$, although it is known that
a distribution of relaxation times is in fact implied for a more adequate
description of the glass transition. We can, however, suppose that
$\tau$ is the average upon all the $\tau_{i}$ implied in the distribution.
With the first assumption, this implies that enthalpy and volume evolve
along time following the same $\tau$ after temperature or pressure
disturbances. 
\item Eventually, we suppose that the glass transition can be well depicted
using linear physics of irreversible processes. This is the case if
temperature or pressure time rates imposed on the system are so slow
that the system never departs too far from equilibrium. This latter
assumption leads up to important mathematical consequences:\end{itemize}
\begin{enumerate}
\item -affinity and the time rate of the order parameter are proportional:\begin{equation}
\frac{d\xi}{dt}\thicksim L\frac{A}{T}\label{eq:Onsager}\end{equation}
 where $L$ is the so-called Onsager coefficient. It is sometimes
discussed in the literature that such a relation can hold even for
high departures from equilibrium. For example, the Fourier law remains
valid even upon high temperature gradients. Prigogine and co-workers
have experimentally validate such relations in the case of particular
chemical reactions where the assumption comes back to neglect $A$
as regard to $RT$ ($R$ is the ideal gas constant) \cite{prigo2}.
Now, discussing the validity of such linear relation for the case
of the glass transition is a problem beyond the scope of this article.
This has been well investigated by Möller and \textit{al.} and Gutzow
and \textit{al.} \cite{gutzow3,gutzow4,gutzow5}. 
\item -the isobaric and isothermal heat and volume advancements of the transformation
(derivatives of the first order with respect to $\xi$ of the enthalpy
and the volume) have always constant values, i.e their values at equilibrium:\begin{equation}
\begin{cases}
\left(\frac{\partial V}{\partial\xi}\right)_{p,T} & \sim\left(\frac{\partial V}{\partial\xi}\right)_{p,T}^{eq}\\
\left(\frac{\partial H}{\partial\xi}\right)_{p,T} & \sim\left(\frac{\partial H}{\partial\xi}\right)_{p,T}^{eq}\end{cases}\end{equation}

\item -the second order derivative of the Gibbs free energy with respect
to $\xi$ has also its value at equilibrium:\begin{equation}
\left(\frac{\partial^{2}G}{\partial\xi^{2}}\right)_{p,T}\sim\left(\frac{\partial^{2}G}{\partial\xi^{2}}\right)_{p,T}^{eq}\end{equation}

\item -the order parameter relaxes towards its equilibrium value following
a simple first order linear differential equation:\begin{equation}
\frac{d\xi}{dt}\sim-\frac{\left(\xi-\xi_{eq}\right)}{\tau}\label{eq:relax ksi}\end{equation}
 The equilibrium value of the order parameter $\xi_{eq}$ is exclusively
and instantaneously driven by the temperature or the pressure. 
\item -The last consequence is certainly the most difficult to justify.
It states that close to equilibrium the affinity is negligible as
compared to the isothermal heat of the transformation:\begin{equation}
A<<\left(\frac{\partial H}{\partial\xi}\right)_{p,T}^{eq}\label{eq:A neglected}\end{equation}
 In Appendix A, we show briefly that this last assumption comes to
neglect the heat of friction due to the entropy production term as
compared to the heat released or supplied due to the advance of the
order parameter at a given temperature. In other words, this means
that the amount of heat involved in the structural change is greater
than the heat produced by internal dissipation. This implies also
that the fictive temperature $T_{f}$ can be, to a first order, close
to the temperature $T$ of the system in the linear regime (but bear
in mind that $dT_{f}\neq dT$ to a first order). 
\end{enumerate}
All these assumptions have been discussed in classical non-equilibrium
thermodynamics books \cite{prigo3,degroot,munster}.

\section{Configurational Prigogine-Defay ratio}

\subsection{Derivation of the configurational Prigogine-Defay ratio for general
out of equilibrium transformations}

A thermodynamic transformation occurs out of equilibrium if the system
of interest is, over a certain time, departed from its state of equilibrium.
This means that the order parameter $\xi$ is deviated from its equilibrium
value $\xi_{eq}$ whereas it evolves along time following the equation
(\ref{eq:relax ksi}). The flux of the order parameter is driven by
the thermodynamic force of the process which corresponds to its conjugated
affinity. Since the affinity is a state function, it can be differentiated
with respect to the three state variables, namely $\left\{ p,T,\xi\right\} $.
This yields the total differential of affinity \cite{prigo}:\begin{equation}
dA=\left(\frac{\partial A}{\partial T}\right)_{p,\xi}dT+\left(\frac{\partial A}{\partial P}\right)_{T,\xi}dp+\left(\frac{\partial A}{\partial\xi}\right)_{p,T}d\xi\label{eq:total diff of A}\end{equation}
 The first and second partial derivatives involved in this equation
can be transformed using Maxwell relations at constant pressure and
temperature respectively. With the help of Eq. (\ref{eq:Berth de donder})
this yields:\begin{equation}
\left(\frac{\partial A}{\partial T}\right)_{p,\xi}=\left(\frac{\partial S}{\partial\xi}\right)_{p,T}=\frac{\left(\frac{\partial H}{\partial\xi}\right)_{p,T}+A}{T}\label{eq:dA/dT}\end{equation}
 \begin{equation}
\left(\frac{\partial A}{\partial P}\right)_{T,\xi}=-\left(\frac{\partial V}{\partial\xi}\right)_{T,p}\label{eq:dA/dp}\end{equation}
 Using the definition of affinity, the third partial derivative is
simply (with an opposite sign) equal to the second order derivative
with respect to $\xi$ of the Gibbs free energy at constant pressure
and temperature:\begin{equation}
\left(\frac{\partial A}{\partial\xi}\right)_{p,T}=-\left(\frac{\partial^{2}G}{\partial\xi^{2}}\right)_{p,T}\end{equation}
 The thermodynamic stability of the system requires that $\left(\frac{\partial^{2}G}{\partial\xi^{2}}\right)_{p,T}$
must be strictly positive. From the total differential of the affinity
(Eq. (\ref{eq:total diff of A})), it is now possible to extract the
isobaric temperature derivative of the order parameter and the isothermal
pressure derivative of the order parameter involved in the three configurational
contributions (Eq. (\ref{eq:deltaCp}), (\ref{eq:deltaKappa}) and
(\ref{eq:delta alpha})):\begin{equation}
\left(\frac{d\xi}{dT}\right)_{p}=\frac{A+\left(\frac{\partial H}{\partial\xi}\right)_{p,T}-T\left(\frac{dA}{dT}\right)_{p}}{T\left(\frac{\partial^{2}G}{\partial\xi^{2}}\right)_{p,T}}\label{eq:dksidT}\end{equation}
 \begin{equation}
\left(\frac{d\xi}{dp}\right)_{T}=-\frac{\left[\left(\frac{\partial V}{\partial\xi}\right)_{T,p}+\left(\frac{dA}{dp}\right)_{T}\right]}{\left(\frac{\partial^{2}G}{\partial\xi^{2}}\right)_{p,T}}\label{eq:dksidp}\end{equation}
 These two temperature and pressure derivatives reveal the affinity
production due to temperature perturbation under isobaric conditions
and due to pressure perturbation under isothermal conditions respectively,
which are the signatures of irreversible processes taking place during
such perturbations. Now, from these three configurational contributions
(Eq. (\ref{eq:deltaCp}), (\ref{eq:deltaKappa}) and (\ref{eq:delta alpha})):\begin{equation}
\triangle C_{p}=C_{p}^{conf}\end{equation}
 \begin{equation}
\triangle\kappa_{T}=\kappa_{T}^{conf}\end{equation}
 \begin{equation}
\triangle\alpha_{p}=\alpha_{p}^{conf}\end{equation}
 a configurational Prigogine-Defay ratio can be derived: \begin{equation}
\Pi^{conf}=\frac{1}{VT}\frac{\triangle C_{p}\triangle\kappa_{T}}{\left(\triangle\alpha\right)^{2}}=-\frac{1}{T}\frac{\left(\frac{\partial H}{\partial\xi}\right)_{p,T}}{\left(\frac{\partial V}{\partial\xi}\right)_{T,p}}\frac{\left(\frac{d\xi}{dp}\right)_{T}}{\left(\frac{d\xi}{dT}\right)_{p}}\end{equation}
 Here the sign {}``$\Delta$'' signifies a difference between thermodynamic
coefficients corresponding to the effective state of the system measured
during the glass transition and those of the corresponding glassy
state (iso-order-parameter state). With Eq. (\ref{eq:dksidT}) and
(\ref{eq:dksidp}) the CPD ratio can be expressed as the following:\begin{equation}
\Pi^{conf}=\frac{\left(\frac{\partial H}{\partial\xi}\right)_{p,T}\left[\left(\frac{\partial V}{\partial\xi}\right)_{T,p}+\left(\frac{dA}{dp}\right)_{T}\right]}{\left(\frac{\partial V}{\partial\xi}\right)_{T,p}\left[A+\left(\frac{\partial H}{\partial\xi}\right)_{p,T}-T\left(\frac{dA}{dT}\right)_{p}\right]}\label{eq:P.D. ratio general}\end{equation}
 This is the general expression of the CPD ratio during a non-equilibrium
glass transition. Some comments are necessary at this point. Firstly,
the CPD ratio is defined over the whole transformation interval (temperature
and pressure intervals) within which the glass transition takes place,
and not only at a predetermined temperature such as $T_{g}$ or a
characteristic pressure within the glass pressure range. This remark
has already been outlined by Schmelzer and Gutzow \cite{gutzow2,gutzow3}.
Secondly, due to the time dependence of affinity (and its derivatives),
$\Pi^{conf}$ can be time dependent. However, time dependent quantities
in Eq. (\ref{eq:P.D. ratio general}) can vary in such a way that
the ratio remains fixed and time independent. The notion of time dependent
PD ratio has already been envisaged by Havlicek \cite{havlicek1,havlicek2}.
Let us, for closing this part, indicate that the assumptions previously
mentioned have not been used for the moment. Now, from the general
expression of the CPD ratio, different cases can be envisaged.

\subsubsection{Ideal glass transition: usual Prigogine-Defay ratio for an equilibrium
thermodynamic transformation}

A transformation which takes place at thermodynamic equilibrium is
a transformation for which the control parameters (pressure or temperature)
vary so slowly that the system has enough time to restore its equilibrium
state. In other words, when $\xi_{eq}(p,T)$ moves, $\xi$ follows
it instantaneously, or in practises, $\xi$ moves more rapidly than
the variations of the control parameters. Consequently, it is impossible
to observe an equilibrium glass transition. If it would be the case,
the liquid must remain a super-cooled liquid during all the pressure
or temperature courses until the liquid becomes a solid. The heat
capacity of the super-cooled liquid can be extrapolated to a characteristic
temperature for which the liquid and glassy states are the same. This
temperature is the Kauzman temperature, $T_{K}$, for which the relaxation
time is infinite. To the best of our knowledge, nobody has never been
able to perform experiments in order to approach sufficiently close
to this characteristic temperature. As of today, it is unknown whether
or not at this temperature a phase transition occurs, or whether there
a discontinuity in the derivatives of the Gibbs free energy. During
a real glass transition, eventually, the relaxation time of the order
parameter becomes so high that it is impossible for the system to
adapt its structure to the new temperature and pressure, respectively,
in the experimental time scale. Thus, a glass transition occurs outside
thermodynamic equilibrium. Anyway, let us suppose for instance that
such an ideal glass transition is possible at the limit of infinitely
slow temperature and pressure rates: $dT/dt\rightarrow0$ and $dp/dt\rightarrow0$.
In that case, the three configurational contributions $\triangle C_{p}$,
$\triangle\kappa_{T}$ and $\triangle\alpha_{p}$ are at their maximum
and always equal to the difference between the equilibrium liquid
and the glass at each temperature and pressure (the glass being the
state of matter obtained for infinitely fast temperature and pressure
rates: $dT/dt\rightarrow\infty$ and $dp/dt\rightarrow\infty$). On
Figure 1, a schematic diagram shows how the heat capacity behaves
during different thermodynamic transformations, i.e during an equilibrium
transformation (liquid), during a perfect non-equilibrium transition
(ideal glass) and during a real glass transition (viscous glass).
At this point, one can ask whether measuring such differences in the
thermodynamic coefficients does make sense for the study of a real
glass transition. Indeed, even in the presence of a real glass transition
(for which $dT/dt$ takes a defined value different from zero), if
the difference between the super-cooled liquid and the glass is taken
at $T_{g}$ by extrapolation of the heat capacity of the liquid and
the glass at this temperature, this comes back to take $\triangle C_{p}^{eq}$
for determining the PD ratio. Anyway, we can calculate these maximal
configurational contributions by means of the total differential of
affinity in putting $A=0$ and $dA=0$. In this situation, temperature
and pressure derivatives of the equilibrium value of the order parameter
become (Eq. \ref{eq:dksidT},\ref{eq:dksidp}): \begin{equation}
\left(\frac{d\xi}{dT}\right)_{p}^{eq}=\frac{\left(\frac{\partial H}{\partial\xi}\right)_{p,T}^{eq}}{T\left(\frac{\partial^{2}G}{\partial\xi^{2}}\right)_{p,T}^{eq}}\end{equation}
 \begin{equation}
\left(\frac{d\xi}{dp}\right)_{T}^{eq}=-\frac{\left(\frac{\partial V}{\partial\xi}\right)_{T,p}^{eq}}{\left(\frac{\partial^{2}G}{\partial\xi^{2}}\right)_{p,T}^{eq}}\end{equation}
 The equilibrium configurational contributions (Eq. \ref{eq:deltaCp},
\ref{eq:deltaKappa}, \ref{eq:delta alpha}) are as follows:\begin{equation}
\triangle C_{p}^{eq}=\frac{\left[\left(\frac{\partial H}{\partial\xi}\right)_{p,T}^{eq}\right]^{2}}{T\left(\frac{\partial^{2}G}{\partial\xi^{2}}\right)_{p,T}^{eq}}\label{eq:deltaCp eq}\end{equation}
 \begin{equation}
\triangle\kappa_{T}^{eq}=\frac{\left[\left(\frac{\partial V}{\partial\xi}\right)_{T,p}^{eq}\right]^{2}}{V\left(\frac{\partial^{2}G}{\partial\xi^{2}}\right)_{p,T}^{eq}}\label{eq:deltaKappa eq}\end{equation}
 \begin{equation}
\triangle\alpha_{p}^{eq}=\frac{\left(\frac{\partial V}{\partial\xi}\right)_{T,p}^{eq}\left(\frac{\partial H}{\partial\xi}\right)_{p,T}^{eq}}{VT\left(\frac{\partial^{2}G}{\partial\xi^{2}}\right)_{p,T}^{eq}}\label{eq:delta alpha eq}\end{equation}
 We can compute these three equilibrium contributions in the formula
of the PD ratio (or simply in putting $A=0$ and $dA=0$ in Eq. (\ref{eq:P.D. ratio general}))
in order to obtain a PD ratio, in its usual meaning, which is strictly
equal to unity. The conclusion is that, under the single order parameter
assumption, during an ideal glass transition the PD ratio is equal
to unity like in the case of a second order phase transition. During
this equilibrium transformation, the PD ratio is measured over a given
temperature or pressure interval and not only at a critical temperature
or pressure. Now a pertinent question could be to know if, under the
single order parameter assumption, the non-unity of the PD ratio could
be due to the real out of equilibrium nature of the glass transition.

\subsubsection{Isoaffine glass transition}

The isoaffine transformation is a very particular case of non-equilibrium
transformation. The fact that this type of transformation can well-describe
a glass transition is, in our point of view, questionable. Indeed,
it must be very improbable that all the partial derivatives in the
equation (\ref{eq:total diff of A}) are connected in such a manner
that $dA$ equals zero at any time during the transformation while
$A$ is different from zero. Moreover, since we know that the glass
transition is a transformation starting from an equilibrium state
(super-cooled liquid) for which $A=0$ to a glassy state for which
$A\neq0$, then the isoaffine glass transition is, in fact, impossible.
Nevertheless, it seems that this has been the way which was initially
followed by Prigogine and Defay, and that this way was also recently
reconsidered by Schmelzer and Gutzow \cite{prigo,gutzow2}. A critical
lecture of the famous paper of Davies and Jones (Ref. \cite{davies})
reveals that most of the mathematical thermodynamic derivations are
also made under the assumption of iso-affinity. Indeed, in making
$dA=0$ in equation (\ref{eq:P.D. ratio general}) then the expression
of the PD ratio derived by Schmelzer and Gutzow is found:\begin{equation}
\Pi^{conf}=\frac{\left(\frac{\partial H}{\partial\xi}\right)_{p,T}}{A+\left(\frac{\partial H}{\partial\xi}\right)_{p,T}}\end{equation}
 Under the assumption of linearity (close to equilibrium transformation)
then for the isoaffine glass transition, $\Pi^{conf}$ is very close
to unity because in this case the inequality $A<<\left(\frac{\partial H}{\partial\xi}\right)_{p,T}$
holds (see Eq. \ref{eq:A neglected}, see also more precision on this
assumption in Appendix A).

\subsubsection{A three dimensional diagram of the glass transition}

The state of a material out of thermodynamic equilibrium is given
by three independent thermodynamic variables $(p,T,A)$. Thus the
state of a glassy material can be represented in a diagram with three
axes representing the values of $p,T$ and $A$ as in Figure 2. The
order parameter is a function of these three variables alone: $\xi=\xi(p,T,A)$
while at equilibrium, the equilibrium order parameter $\xi_{eq}=\xi_{eq}(p,T,A=0)$
is fixed by the condition $A=0$ and is a function of $p,T$ only.
Upon cooling at constant pressure $p=p_{0}$ but with different temperature
rates $\gamma_{1}$ and $\gamma_{2}$, the system is driven out of
equilibrium by the temperature variation rate and enters into the
slowly relaxing states at different temperatures $T_{a}$ and $T_{b}>T_{a}$.
Upon further cooling, the system enters a true glassy state with a
fixed order parameter $\xi_{a}$ and $\xi_{b}$. These evolutions
are represented as trajectories (a) and (b) on the Figure 2. Although
glass 1 and glass 2, represented by the point (a) and (b), are at
the same temperature and pressure, it is clear that they are not in
the same states since they have different affinities. Nevertheless
if these two glassy materials were still within the glass transition
interval, they would relax toward the $A=0$ and eventually reach
identical equilibrium states. The fictive temperature concept is also
a measure of the structure of the system like $\xi$ (cf. following
paragraphs). According to the definitions given by Davies and Jones
\cite{davies}, the fictive temperature $T_{f}$ of the materials
can be deduced in the linear regime by the intercept of the tangent
of the trajectories in the $(p_{0},T,A)$ plane with the $(p_{0},T,A=0)$
axis. In the glass transition interval, the fictive temperature is
varying and follows the control temperature although with some delay,
its value at a particular point within the glass transition range
of trajectory (b) has been represented on the diagram of figure 2.
As soon as the system is in a glassy state but still in the linear
regime, the fictive temperature is constant and noted $T_{f}^{'}$
and necessary the trajectory in $(p_{0},T,A)$ plane becomes rectilinear.
The final fictive temperatures ($T_{f}^{'}$) have been marked on
the diagram for trajectories (a) and (b). Next we perform a cooling
of the material under pressure $p_{1}$ at a rate $\gamma_{2}$ as
represented by trajectory (c). We explicitly supposed in the diagram
that $dT_{g}/dp>0$ (where $T_{g}$ is the glass temperature). In
the case of a true first order transition, this would be equivalent
to supposing that the molar volume of the glass is smaller than the
molar volume of the liquid for a positive melting latent heat. In
other words, this means that under pressure for a given temperature
rate, the system is driven out of equilibrium at higher temperature
than under atmospheric pressure \cite{Wondraczek1,Wondraczek2,wondraczek3}.
This can be re-phrased in terms of fictive temperature: for a given
cooling rate, the fictive temperature of a sample cooled under high
pressure will be higher than the fictive temperature of a sample cooled
under atmospheric pressure. This has been recently observed experimentally
\cite{Wondraczek1,Wondraczek2,wondraczek3}. The final fictive temperature
has also been marked for this glass. Finally we perform an experiment
in which the system is maintained in equilibrium $(A=0)$ at constant
temperature $T_{0}$. The pressure is then increased at a given rate
yielding the trajectory (d) pictured on figure 2. If the pressure
variation drives the system out of thermodynamic equilibrium, then
above a vitrifying pressure $p_{g}$, the system enters first a glass
transition range in pressure and if the pressure is further increased,
the system becomes totally glassy and increasing the pressure does
not change significantly the value of the order parameter $\xi$.
So, if the cooling rate of trajectory (c) and the increasing pressure
rate of trajectory (d) are well chosen, it is possible that the two
trajectories intercept as pictured by point (d) in the diagram. A
fictive pressure of the glassy material (d) can be defined in the
same manner as the fictive temperature. From this diagram, it can
be seen that the system can be set into an identical glassy state
by using either temperature variation at constant pressure or pressure
variation at constant temperature. This naturally provides a link
and equivalence between the fictive pressure and fictive temperature
concepts and the order parameter approach introduced by De Donder.
Figure 3 provides an experimental illustration of above arguments.
It shows the dependence of a structural parameter, the relative amount
of tetrahedrally coordinate boron, $BO_{4}$, in a borosilicate glass
as a function of fictive pressure and fictive temperature \cite{Wondraczek1}.
Here, $\left[BO_{4}\right]$ may be taken as related to the order
parameter, and different $\left(T,p\right)$ paths may lead to different
structural states of the final glass. In the following, a formalism
is developed that will help to translate future physical experiments
in the framework of the CPD ratio.

\subsection{Expression of the Configurational Prigogine-Defay ratio as a function
of the isobaric heat capacity and isothermal compressibility coefficient}

From now, the assumptions described in paragraph 2.2 will be used
in order to simplify the general expression (\ref{eq:P.D. ratio general})
of the CPD ratio. This will allow to connect its value to measurable
quantities in a different, novel fashion. Firstly, let us re-formulate
expression (\ref{eq:P.D. ratio general}) in the form of a ratio of
two adimensional quantities that involve only thermal and mechanical
variables, respectively:\begin{equation}
\Pi^{conf}=\frac{\Pi_{calo}}{\Pi_{meca}}\end{equation}
 where we have for the two terms of the ratio:\begin{equation}
\Pi_{calo}=\frac{\left(\frac{\partial H}{\partial\xi}\right)_{p,T}}{\left[A+\left(\frac{\partial H}{\partial\xi}\right)_{p,T}-T\left(\frac{dA}{dT}\right)_{p}\right]}\end{equation}
 \begin{equation}
\Pi_{meca}=\frac{\left(\frac{\partial V}{\partial\xi}\right)_{T,p}}{\left[\left(\frac{\partial V}{\partial\xi}\right)_{T,p}+\left(\frac{dA}{dp}\right)_{T}\right]}\end{equation}
 Using the assumptions of paragraph 2.2, we obtain:\begin{equation}
\Pi_{calo}\sim\frac{\left(\frac{\partial H}{\partial\xi}\right)_{p,T}^{eq}}{\left[\left(\frac{\partial H}{\partial\xi}\right)_{p,T}^{eq}-T\left(\frac{dA}{dT}\right)_{p}^{eq}\right]}\end{equation}
 \begin{equation}
\Pi_{meca}\sim\frac{\left(\frac{\partial V}{\partial\xi}\right)_{T,p}^{eq}}{\left[\left(\frac{\partial V}{\partial\xi}\right)_{T,p}^{eq}+\left(\frac{dA}{dp}\right)_{T}^{eq}\right]}\end{equation}
 It could be easily demonstrated that the two previous quantities
can be expressed as a function of the heat capacity only for $\Pi_{calo}$
and as a function of the compressibility only for $\Pi_{meca}$ (see
Appendix B):\begin{equation}
\Pi_{calo}\sim\frac{\Delta C_{p}^{eq}}{\Delta C_{p}}=\frac{\Delta C_{p}^{eq}}{C_{p}^{conf}}\end{equation}
 \begin{equation}
\Pi_{meca}\sim\frac{\triangle\kappa_{T}^{eq}}{\triangle\kappa_{T}}=\frac{\triangle\kappa_{T}^{eq}}{\kappa_{T}^{conf}}\end{equation}
 $\Delta C_{p}^{eq}$ is the equilibrium contribution as classically
involved in the usual PD ratio although $\Delta C_{p}$ is the difference
between the effectively measured heat capacity during the glass transition
(viscous state) and the glassy state. The same arguments hold for
the measured isothermal compressibility. Thus, the CPD ratio now depends
only on the isobaric heat capacity and the isothermal compressibility
(the isobaric thermal expansion coefficient does not now play a role
anymore):\begin{equation}
\Pi^{conf}\sim\frac{\Delta C_{p}^{eq}}{\Delta C_{p}}\frac{\triangle\kappa_{T}}{\triangle\kappa_{T}^{eq}}\end{equation}
 Let us remark that $\Pi_{calo}$is what is known in the glass literature
as the inverse of the normalised heat capacity which is equal to unity
only when the system is in the liquid state. In letting time to the
material to adjust its configuration at each pressure and temperature,
respectively, we observe that $\Pi_{calo}$ and $\Pi_{meca}$ tend
directly toward unity. The question is whether this attainment holds
on such a manner that $\Pi^{conf}$ remains equal to one at each instant.
This could be envisaged in developing the latter expression of $\Pi^{conf}$
using the fact that $\triangle C_{p}$ and $\triangle\kappa_{T}$
involved temperature rate and pressure rate, respectively:\begin{equation}
\triangle C_{p}=\left(\frac{\partial H}{\partial\xi}\right)_{p,T}\frac{d\xi/dt}{dT/dt}\sim\left(\frac{\partial H}{\partial\xi}\right)_{p,T}^{eq}\frac{d\xi/dt}{dT/dt}\end{equation}
 \begin{equation}
\triangle\kappa_{T}=-\frac{1}{V}\left(\frac{\partial V}{\partial\xi}\right)_{T,p}\frac{d\xi/dt}{dp/dt}\sim-\frac{1}{V}\left(\frac{\partial V}{\partial\xi}\right)_{T,p}^{eq}\frac{d\xi/dt}{dp/dt}\end{equation}
 With these previous equations, in assuming that near equilibrium,
a temperature variation under isobaric condition induces the same
order parameter disequilibrium as does a pressure variation under
isothermal condition, with the help of (Eq. (\ref{eq:deltaCp eq})
and (\ref{eq:deltaKappa eq})) we obtain for $\Pi^{conf}$ the following
value:\begin{equation}
\Pi^{conf}\sim-\frac{1}{T}\frac{\left(\frac{\partial H}{\partial\xi}\right)_{p,T}^{eq}}{\left(\frac{\partial V}{\partial\xi}\right)_{T,p}^{eq}}\frac{dT/dt}{dp/dt}\label{eq:Pi conf relax}\end{equation}
 This provides a condition for the unity of the CPD ratio under relaxation
involving mechanical and thermal variables and temperature and pressure
time rates:\begin{equation}
\frac{dT/dt}{dp/dt}=-T\frac{\left(\frac{\partial V}{\partial\xi}\right)_{T,p}^{eq}}{\left(\frac{\partial H}{\partial\xi}\right)_{p,T}^{eq}}\label{eq:requirement}\end{equation}
 This is a prerequisite for experiments in order to obtain a CPD ratio
equal to unity. The experimental challenge is to be able to follow
this requirement for being at the same point in the three dimensional
diagrams of figures 2 and 3 after temperature and (or) pressure perturbations.
Noteworthy, as mentioned before, if enthalpy and volume relax towards
equilibrium with different relaxation times for equivalent thermodynamic
perturbations, the ratio of the two relaxation times ($\tau_{H}$
and $\tau_{V}$) must appear in the previous equation. In other words,
if thermal and mechanical variables are driven by different order
parameters, then experimentally, we must adapt the rate of variations
of $T$ in one experiment and the rate of variation of $P$ in another
one in order to fulfil the unity condition for the CPD ratio. Wondraczek
and colleagues have performed experiments in order to probe such $P,$$T$
influences on the structure of glasses \cite{Wondraczek1,Wondraczek2,wondraczek3}.

\subsection{Expression of the configurational Prigogine-Defay ratio as a function
of the fictive temperature and the fictive pressure}

Tool has originally defined a concept, called the fictive temperature,
adapted to a phenomenological description of calorimetric experiments
on glasses \cite{tool1,tool2}. In short, this quantity represents
the temperature of the slow relaxing modes during a glass transition.
Moynihan and co-workers have used the fictive temperature concept
for a description of the glass transition and ageing processes, and
have connected it to heat capacity data \cite{moy1,moy2,moy3}. By
means of a systematic conformity between the {}``two temperatures
thermodynamics'' developed by Nieuwenhuizen and the order parameter
approach of De Donder, we have recently demonstrated that the fictive
temperature of Tool can in fact represent a real thermodynamic temperature
for systems undergoing irreversible processes \cite{garden1}. For
instance, the fictive temperature of Tool is equivalent to the effective
temperature of Nieuwenhuizen. Davies and Jones have however already
established such a direct link between the fictive temperature and
the order parameter \cite{davies}. In the same way, they also have
defined a fictive pressure. Close to equilibrium, the affinity can
be developed as a function of the fictive temperature departure from
the classical one (under isobaric condition) and also as the function
of the fictive pressure departure from the pressure (under isothermal
condition):\begin{equation}
A\thicksim(T-T_{f})\left(\frac{\partial A}{\partial T}\right)_{p,\xi}^{eq}\end{equation}
 \begin{equation}
A\thicksim(p-p_{f})\left(\frac{\partial A}{\partial p}\right)_{T,\xi}^{eq}\end{equation}
 With equations (\ref{eq:dA/dT}) and (\ref{eq:dA/dp}) and neglecting
the affinity as compared to the heat of the transformation, we obtain:\begin{equation}
A\thicksim\left(\frac{\partial H}{\partial\xi}\right)_{p,T}^{eq}\frac{(T-T_{f})}{T}\end{equation}
 \begin{equation}
A\sim-\left(\frac{\partial V}{\partial\xi}\right)_{T,p}^{eq}(p-p_{f})\end{equation}
 Using linear relation (\ref{eq:Onsager}) this yields to a new expression
for $\Pi^{conf}$ from Eq. (\ref{eq:Pi conf relax}):\begin{equation}
\Pi^{conf}\sim\frac{(p-p_{f})}{(T-T_{f})}\frac{dT/dt}{dp/dt}\label{eq:Piconf fict1}\end{equation}
 Using the well-know differential equation of Tool and a similar one
for the fictive pressure in the linear regime we have:\begin{equation}
\frac{dT_{f}}{dt}=\frac{(T-T_{f})}{\tau}\end{equation}
 \begin{equation}
\frac{dp_{f}}{dt}=\frac{(p-p_{f})}{\tau}\end{equation}
 When $T$ or $p$ are stopped, these two linear differential equations
give an exponential decay for the fictive temperature and the fictive
pressure which are linked to enthalpy and volume relaxation. As we
are placed in the assumption of one single order parameter ($\tau$
is the same in the two equations), this leads to a simple expression
for the CPD ratio:\begin{equation}
\Pi^{conf}\thicksim\frac{dp_{f}}{dp}\frac{dT}{dT_{f}}\label{eq:PI conf fictive}\end{equation}
 It is straightforward to remark that a result obtained a long time
ago by Moynihan, connecting the normalised heat capacity with the
temperature derivative of the fictive temperature has been re-found
\cite{moy1,moy3}:\begin{equation}
\Pi_{calo}\sim\frac{\Delta C_{p}^{eq}}{\Delta C_{p}}=\frac{dT}{dT_{f}}\end{equation}
 Here $\Delta C_{p}^{eq}$ is taken at $T$ while in the original
derivation of Moynihan $\Delta C_{p}^{eq}$ is taken at $T_{f}$ in
the normalised heat capacity expression. The two ways are equivalent
upon the linear assumption because $T\sim T_{f}$ (see Appendix A
for details). The conclusion of this paragraph is that the CPD ratio
can involve only fictive temperature and fictive pressure along with
the classical temperature and pressure. Notwithstanding its simplicity,
this sort of connection can be a good way to establish the link between
this theoretical approach and physical experiments. Relation (\ref{eq:PI conf fictive})
can however indicate to us how, firstly the CPD ratio must be equal
to unity or not, and secondly, how the classical PD ratio, which is
the long-term average of $\Pi^{conf}$, can also be equal to unity
or not. This is briefly discussed now.

\section{On the unity of the Configurational Prigogine-Defay ratio}

At infinitely long time-scale, i.e for infinitely slow cooling rate
or pressure variations, the system has enough time to restore its
internal equilibrium. It behaves like a liquid. The fictive temperature
and pressure vary exactly like their corresponding temperature and
pressure driving variables. Eq. (\ref{eq:PI conf fictive}) states
that $\Pi^{conf}$ is equal to unity. But as already mentioned, under
these circumstances, no glass transition does\ occur. In real situations,
even for very slow variations of the driving variables, there exists
an instant at which the relaxation time of the system $\tau(T,p)$
will reach a value for which the system cannot anymore restore its
long range configuration. From this instant, the experimenter begins
to observe a fall in the measured thermodynamic coefficients ($C_{p}$,
etc..). This fall is not sudden. It occurs over a definite interval
(temperature or pressure). During this collapse, the system looses
its configurational entropy which is observable by a smooth jump in
the heat capacity. This being, there is also relaxational effect which
becomes small due to the importance of the relaxation time. There
is however an irreversible generation of entropy. This entropy gain
due to friction during slow internal configurational readjustments
is generally small. In the linear regime, the entropy production term
is indeed negligible (see Appendix A). This has already been discussed
by Davies and Jones \cite{davies}. From this situation, the fundamental
question arises as to whether the departure of the fictive temperature
from the equilibrium temperature of the system reaches a value in
such a fashion that the fictive pressure attains a value different
from the pressure of the system such that the Eq. (\ref{eq:PI conf fictive})
leads to unity for $\Pi^{conf}$. Indeed, from Eq. (\ref{eq:Piconf fict1}),
considering that only one single order parameter is implied, $\Pi^{conf}$
is unity if the scanning temperature or pressure rates (under isobaric
and isothermal conditions respectively) are in proportion with the
departure of the fictive temperature and pressure respectively from
their driving variables:\begin{equation}
\frac{dT/dt}{dp/dt}=\frac{(T-T_{f})}{(p-p_{f})}\end{equation}
 Here, the requirement obtained in paragraph 3.2 (Eq. (\ref{eq:requirement}))
is found as a function of the fictive temperature and the fictive
pressure directly, which are measurable quantities. The discussion
made after the Eq. (\ref{eq:requirement}) is still valuable. The
fact that the CPD ratio should be equal to unity only if this type
of statement is fulfilled is an argument in favour of researchers
(see for example McKenna \cite{mckenna1}) who argue that the departure
from unity for $\Pi$ remarked in the literature is probably due to
a erroneous consideration of the different histories in the measurement
of the three thermodynamic coefficients $C_{p}$, $\kappa_{T}$ and
$\alpha_{p}$. However, the statement that {}``$\Pi$ different from
unity'' is due to the presence of more than one single order parameter
is also true as we have shown that in this paper. But, to our point
of view, if we suppose that {}``$\Pi$ different from unity'' is
probably due to a bad consideration of the history of a glass sample
in the measurement of the thermodynamic coefficients, then {}``$\Pi$
different from unity'' is not a definitive proof in favour of a multi-order
parameters description of the glass transition.

\section{Conclusion}

In this paper, we have shown that, when the configurational parts
of the three thermodynamic coefficients involved in the expression
of the PD ratio are considered, a general expression can be derived
for this ratio. We have called this general expression, the configurational
PD ratio. The classical PD ratio usually discussed in the literature,
and experimentally measured by extrapolation of the properties of
the super-cooled liquid and the glass at $T_{g}$, is a limiting case
of this expression for infinitely high time scales of observation
(or identically for infinitely slow temperature and pressure time
variations), i.e when only thermodynamic equilibrium is considered.
We have questioned the pertinence of the classical expression of the
PD ratio for the characterization of the glass transition since it
is known that the glass transition is by essence an out of equilibrium
transformation. Since when classical PD ratios are measured, the obtained
values are generally different from unity, we have discussed that
it could be due to the out of equilibrium nature of the glass transition.
Indeed, in a real glass transition the configurational parts of the
three coefficients $C_{p}$, $\kappa_{T}$ and $\alpha_{p}$ vanish
gradually upon temperature decrease or by varying the pressure. Some
classical and some more recent studies have tackled this issue in
discussing the PD ratio in terms of difference of the thermodynamic
coefficients between a state at constant affinity and a state at constant
order parameter. We have also criticised this point of view, because
during the out of equilibrium phase transformation there is a permanent
creation of affinity and the isoaffine case is not typical for the
glass transition. If the out of equilibrium character of the glass
transition must be correctly taken into account then we must construct
a PD ratio with thermodynamic coefficients effectively recorded during
the glass transition, i.e by taking into account a difference between
an out of equilibrium state during the transition (measured $C_{p}$,
$\kappa_{T}$ and $\alpha_{p}$) and a state at constant order parameter
(the glass: $C_{p,\xi}$, $\kappa_{T,\xi}$ and $\alpha_{p,\xi}$).
When taking into account such time dependent variations of the three
thermodynamic coefficients, under the single order parameter assumption
and under the assumption of identical thermodynamic histories, the
unity for the CPD ratio is always confirmed. Our conclusion at this
level joins that which can be found in the literature on this topics,
i.e, that the non unity of the classical PD ratio is either due to
the presence of several order parameters for a complete description
of the glass transition, or because in most of the cases $\triangle C_{p}$,
$\triangle\kappa_{T}$ and $\triangle\alpha_{p}$ are measured on
sample for which thermodynamic history have not seriously be taken
into account \cite{mckenna1}. For the moment the debate is not closed.
Under linearity conditions, we have also given an expression for the
CPD ratio where the number of relevant coefficients have been reduced
from three to two, and where only normalised isobaric heat capacity
and normalised isothermal compressibility play a role. From this last
expression the CPD ratio can be expressed as a function of not only
the temperature and the pressure but also as a function of the fictive
temperature and the fictive pressure. All these properties are experimentally
accessible, and a future challenge will be to correlate the present
formalism to experimental situations.

\bibliographystyle{unsrt} 
\clearpage{}

\begin{figure}
\begin{centering}
\includegraphics[width=1\columnwidth]{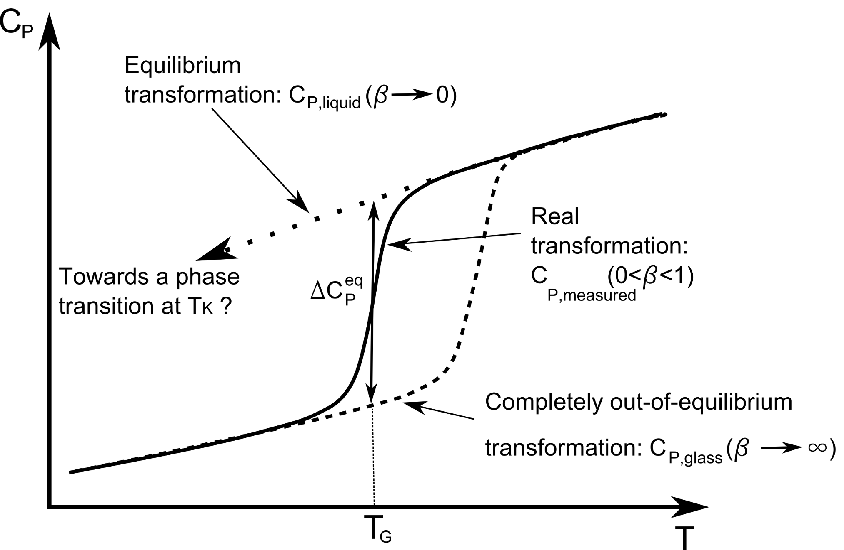}
\par\end{centering}

\caption{On this schematic graph $C_{P}$ as a function of temperature upon
cooling ($\beta=dT/dt$) is shown in different situations. An equilibrium
transformation occurs when the cooling ramp is so slow that equilibrium
is maintained all the time, i.e the measured heat capacity is that
of a super-cooled liquid. A completely non-equilibrium transformation
occurs when the cooling ramp is infinitely fast. The system becomes
instantaneously frozen, i.e the measured heat capacity is that of
a glass. A real transformation takes place when a liquid is cooled
following a determined scanning rate where it undergoes a measurable
jump in the heat capacity. The difference $\triangle C_{p}^{eq}$
which is computed in the classical PD ratio is taken between the hypothetically
equilibrium and completely non-equilibrium state taken at the glass
temperature $T_{g}$ of the real glass transition.}

\end{figure}

\clearpage{}

\begin{figure}
\begin{centering}
\includegraphics[width=1\columnwidth]{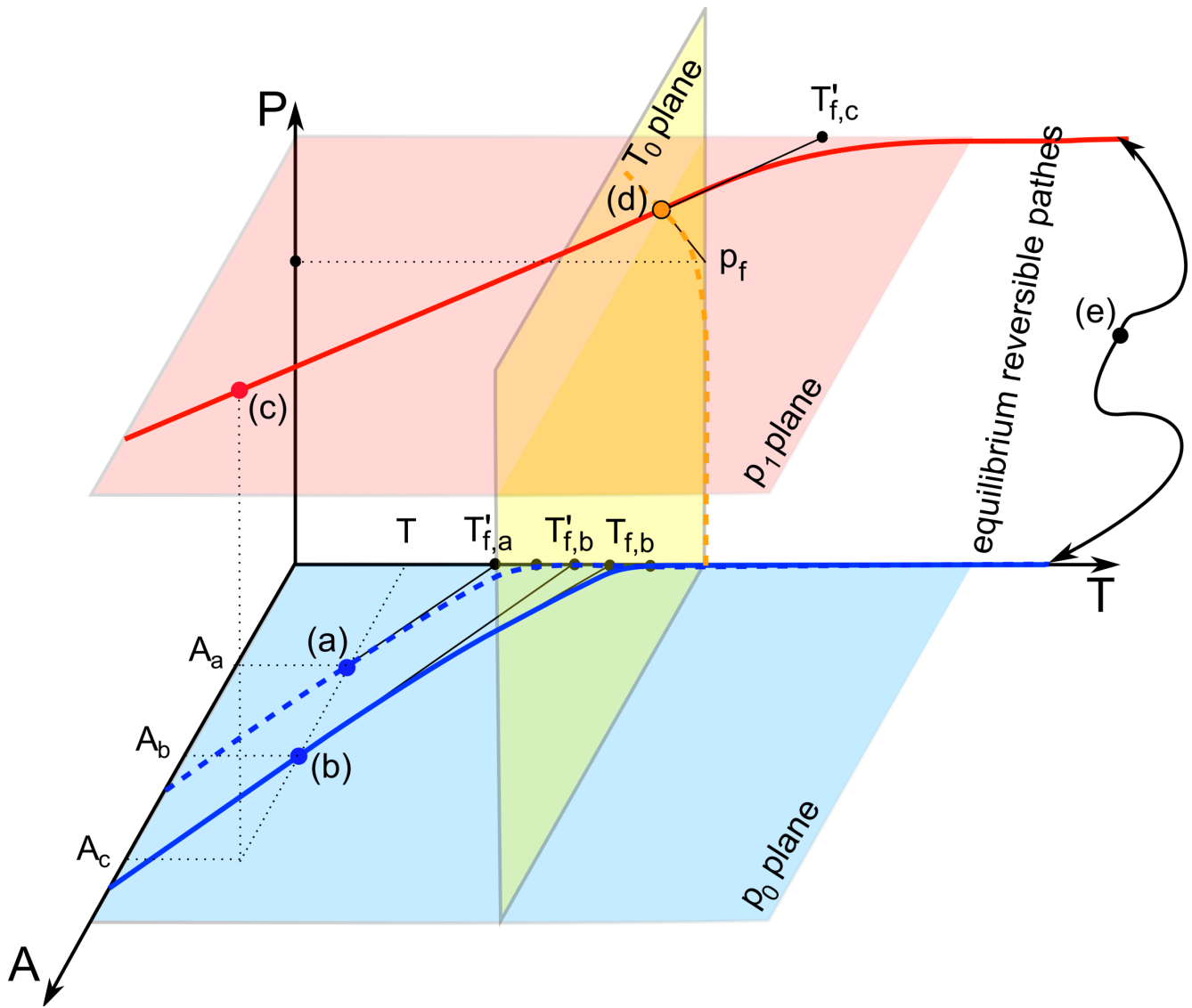}
\par\end{centering}

\caption{Diagram in the $(p,T,A)$ space representing the state of an out of
equilibrium system. Different hypothetical trajectories yielding glasses
are represented: (a) $p=p_{0};\;\frac{dT}{dt}=\gamma_{1}$; (b) $p=p_{0};\;\frac{dT}{dt}=\gamma_{2}>\gamma_{1}$;
(c) $p=p_{1}>p_{0};\frac{dT}{dt}=\gamma_{1}$; (d) $T=T_{0};\;\frac{dp}{dT}\neq0$.
The corresponding glass states are marked by the letters (a), (b),
(c) and (d). The equilibrium state from which the glasses are formed
is marked by the letter (e).are connected by equilibrium (reversible)
trajectories in the $(p,T,A=0)$ plane.}

\end{figure}

\clearpage{}

\begin{figure}
\begin{centering}
\includegraphics[width=1\columnwidth]{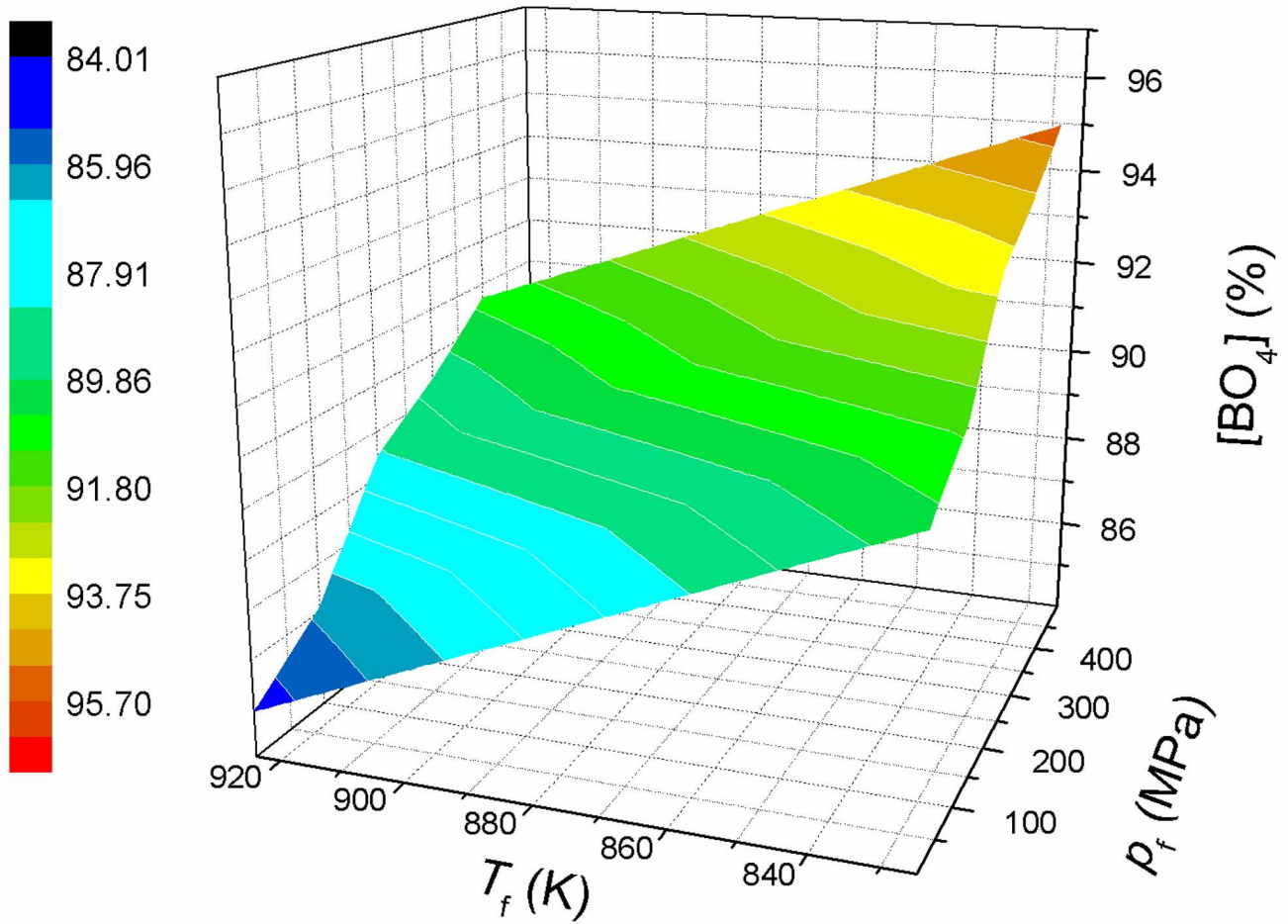}
\par\end{centering}

\caption{$\left\{ P_{f},T_{f},[BO_{4}]\right\} $tri-dimensional graph showing
the dependence of a structural parameter, the relative amount of tetrahedrally
coordinate boron, $BO_{4}$, in a borosilicate glass as a function
of fictive pressure and fictive temperature \cite{Wondraczek1}. The
concentration of boron $\left[BO_{4}\right]$ is the structural order
parameter in these experiments.}

\end{figure}

\clearpage{}

\part*{Appendix A:}

\title{Appendix A}

One assumes that sufficiently close to equilibrium the affinity obeys
the inequality:\begin{equation}
A<<\left(\frac{\partial H}{\partial\xi}\right)_{p,T}^{eq}\label{eq:A neglect}\end{equation}
 Dividing by $T$ and multiplying by $d\xi/dt$ the two members yields:\begin{equation}
\frac{A}{T}\frac{d\xi}{dt}<<\frac{\left(\frac{\partial H}{\partial\xi}\right)_{p,T}^{eq}\frac{d\xi}{dt}}{T}\end{equation}
 The left-hand-side is the so-called entropy production term and the
right-hand-side the configurational entropy rate:\begin{equation}
\frac{d_{i}S}{dt}<<\frac{dS^{conf}}{dt}\end{equation}
 The configurational entropy rate is defined thanks to the configurational
heat capacity such as follows:\begin{equation}
\frac{dS^{conf}}{dt}=\frac{C_{p}^{conf}\times\frac{dT}{dt}}{T}\end{equation}
 Thus neglecting the affinity as compared to the heat of reaction
at constant temperature and pressure is equivalent to neglect the
entropy production as compared to the configurational entropy change.

Davies and Jones have shown that close to equilibrium the affinity
can be written simply as follows \cite{davies}:\begin{equation}
A\thicksim(T-T_{f})\left(\frac{\partial A}{\partial T}\right)_{p,\xi}^{eq}\end{equation}
 This comes simply from a Taylor series expansion with an arrest to
a first order in the departure of the fictive temperature to the temperature.
Using Maxwell relation and the Berthelot-De Donder formula (Eq. \ref{eq:Berth de donder}),
this last equation becomes:\begin{equation}
A\thicksim\frac{(T-T_{f})}{T}\left[\left(\frac{\partial H}{\partial\xi}\right)_{p,T}^{eq}+A\right]\label{eq:A non neglect}\end{equation}
 Let us first observe that if the inequality (\ref{eq:A neglect})
is fulfilled then the affinity can be written:\begin{equation}
A\thicksim\frac{(T-T_{f})}{T}\left(\frac{\partial H}{\partial\xi}\right)_{p,T}^{eq}\label{eq:A less exact}\end{equation}
 Secondly, if we develop further the previous equation (\ref{eq:A non neglect}),
without using the inequality (\ref{eq:A neglect}), we arrive easily
to the following equation:\begin{equation}
A\thicksim\frac{(T-T_{f})}{T_{f}}\left(\frac{\partial H}{\partial\xi}\right)_{p,T}^{eq}\label{eq:A more exact}\end{equation}
 In comparing the two previous equations, we conclude that neglecting
the affinity before the heat of reaction in the linear regime is also
equivalent to state that the fictive temperature remains close to
the real temperature of the system to a first order:\begin{equation}
T_{f}\sim T\end{equation}
 Let us be aware, such as shown in the paper, that it is not true
for the first order derivative of these two temperatures even to a
first order:\begin{equation}
dT_{f}\neq dT\end{equation}
 Knowing this fact, we can start from the expression of the PD ratio
recently derived by Schmelzer and Gutzow for isoaffine case \cite{gutzow2}:\begin{equation}
\Pi_{A=cste}^{conf}=\frac{\left(\frac{\partial H}{\partial\xi}\right)_{p,T}}{A+\left(\frac{\partial H}{\partial\xi}\right)_{p,T}}\end{equation}
 And using the equality (\ref{eq:A more exact}) we obtain easily
that:\begin{equation}
\Pi_{A=cste}^{conf}\sim\frac{T_{f}}{T}\end{equation}
 which, as demonstrated, in the linear regime is close to the unity.
Such an expression for the PD ratio has already been obtained by Ottinger
by an other way \cite{ottinger}. If we start mostly from the equality
(\ref{eq:A less exact}), that we have derived recently \cite{garden1},
we obtain for the isoaffine PD ratio the other expression:\begin{equation}
\Pi_{A=cste}^{conf}\sim\frac{T}{2T-T_{f}}\end{equation}
 which is also very close to the unity in the linear regime.

A last remark can be made: in using the Eq. (\ref{eq:A more exact})
in place of Eq. (\ref{eq:A less exact}) in the Eq. (\ref{eq:Pi conf relax})
then, a more exact expression is found for the expression (\ref{eq:PI conf fictive})
which becomes now:\begin{equation}
\Pi^{conf}\thicksim\frac{T_{f}}{T}\frac{dp_{f}}{dp}\frac{dT}{dT_{f}}=\frac{dp_{f}}{dp}\frac{dlnT}{dlnT_{f}}\end{equation}
 Upon the assumption used, this is equivalent to the expression (\ref{eq:PI conf fictive}).

\clearpage{}

\part*{Appendix B:}

\title{Appendix B}

Firstly, let us rewrite the exact expressions of $\Pi_{calo}$ and
$\Pi_{meca}$:\begin{equation}
\Pi_{calo}=\frac{\left(\frac{\partial H}{\partial\xi}\right)_{p,T}}{\left[A+\left(\frac{\partial H}{\partial\xi}\right)_{p,T}-T\left(\frac{dA}{dT}\right)_{p}\right]}\label{eq:pi calo}\end{equation}
 \begin{equation}
\Pi_{meca}=\frac{\left(\frac{\partial V}{\partial\xi}\right)_{T,p}}{\left[\left(\frac{\partial V}{\partial\xi}\right)_{T,p}+\left(\frac{dA}{dp}\right)_{T}\right]}\label{eq:pi meca}\end{equation}
 For the equation (\ref{eq:pi calo}), in multiplying the numerator
and denominator by the ratio $\left(\frac{\partial H}{\partial\xi}\right)_{p,T}/T\left(\frac{\partial^{2}G}{\partial\xi^{2}}\right)_{p,T}$,
we obtain the expression:\begin{equation}
\Pi_{calo}=\frac{\left[\left(\frac{\partial H}{\partial\xi}\right)_{p,T}\right]^{2}/T\left(\frac{\partial^{2}G}{\partial\xi^{2}}\right)_{p,T}}{\left[A\left(\frac{\partial H}{\partial\xi}\right)_{p,T}/T\left(\frac{\partial^{2}G}{\partial\xi^{2}}\right)_{p,T}+\left(\frac{\partial H}{\partial\xi}\right)_{p,T}^{2}/T\left(\frac{\partial^{2}G}{\partial\xi^{2}}\right)_{p,T}-\left(\frac{dA}{dT}\right)\left(\frac{\partial H}{\partial\xi}\right)_{p,T}/\left(\frac{\partial^{2}G}{\partial\xi^{2}}\right)\right]}\end{equation}
 where we recognize for the numerator an expression close to the maximum
contribution of the configurational heat capacity $\triangle C_{p}^{eq}$
(see Eq. \ref{eq:deltaCp eq}). Upon the assumptions used, where it
is admitted that $\left(\frac{\partial H}{\partial\xi}\right)_{p,T}\thicksim\left(\frac{\partial H}{\partial\xi}\right)_{p,T}^{eq}$
and $\left(\frac{\partial^{2}G}{\partial\xi^{2}}\right)_{p,T}\thicksim\left(\frac{\partial^{2}G}{\partial\xi^{2}}\right)_{p,T}^{eq}$
, then it is equivalent to $\triangle C_{p}^{eq}$. For the denominator,
we recognize $\Delta C_{p}$ expressed in function of the affinity
and the temperature derivative of the affinity (see Ref. \cite{garden2}).
Indeed, in using the total differential of the affinity under isobaric
condition, we can easily replaced $\left(d\xi/dT\right)_{p}$ in the
classical expression of the heat capacity:\begin{equation}
C_{p}=\left(\frac{dH}{dT}\right)_{p}=\left(\frac{\partial H}{\partial T}\right)_{p,\xi}+\left(\frac{\partial H}{\partial\xi}\right)_{p,T}\left(\frac{d\xi}{dT}\right)_{p}\end{equation}
 in order to obtain for $\Delta C_{p}=C_{p}-C_{p,\xi}$, which is
the desired expression. In conclusion, close to equilibrium there
is:\begin{equation}
\Pi_{calo}\sim\frac{\Delta C_{p}^{eq}}{\Delta C_{p}}\end{equation}
 Starting from the equation (\ref{eq:pi meca}), the same reasoning
can be done. Let us start from the definition of the compressibility
coefficient:\begin{equation}
\kappa_{T}=-\frac{1}{V}\left(\frac{dV}{dP}\right)_{T}-\frac{1}{V}\left(\frac{\partial V}{\partial\xi}\right)_{T,p}\left(\frac{d\xi}{dp}\right)_{T}\end{equation}
 In using the total differential of the affinity under isothermal
condition we have obtained:\begin{equation}
\left(\frac{d\xi}{dp}\right)_{T}=-\frac{\left[\left(\frac{\partial V}{\partial\xi}\right)_{T,p}+\left(\frac{dA}{dp}\right)_{T}\right]}{\left(\frac{\partial^{2}G}{\partial\xi^{2}}\right)_{p,T}}\end{equation}
 which included in the preceding equality yields to:\begin{equation}
\Delta\kappa_{T}=\frac{\left[\left[\left(\frac{\partial V}{\partial\xi}\right)_{T,p}\right]^{2}+\left(\frac{\partial V}{\partial\xi}\right)_{T,p}\left(\frac{dA}{dp}\right)_{T}\right]}{V\left(\frac{\partial^{2}G}{\partial\xi^{2}}\right)_{p,T}}\end{equation}
 We observe that such expression appears directly if we multiply the
numerator and denominator of the equation (\ref{eq:pi meca}) by $\left(\frac{\partial V}{\partial\xi}\right)_{T,p}/V\left(\frac{\partial^{2}G}{\partial\xi^{2}}\right)_{p,T}$
, and in this case the maximal contribution of the configurational
compressibility coefficient (in the linear regime) appears directly:
\begin{equation}
\triangle\kappa_{T}^{eq}=\frac{\left[\left(\frac{\partial V}{\partial\xi}\right)_{T,p}^{eq}\right]^{2}}{V\left(\frac{\partial^{2}G}{\partial\xi^{2}}\right)_{p,T}^{eq}}\end{equation}
 Consequently:\begin{equation}
\Pi_{meca}\sim\frac{\Delta\kappa_{T}^{eq}}{\Delta\kappa_{T}}\end{equation}

\end{document}